\documentclass[fleqn]{interact}

\usepackage{epstopdf}
\usepackage{subfigure}

\usepackage{float}

\theoremstyle{plain}

\theoremstyle{definition}

\theoremstyle{remark}

\begin{document}


\title{Spherically symmetric and static solutions in $f(R)$ gravity coupled with EM fields}

\author{
\name{R.~A. Hurtado\textsuperscript{a}\thanks{CONTACT R.~A. Hurtado. Email: rahurtadom@unal.edu.co} and R. Arenas\textsuperscript{a}}
\affil{\textsuperscript{a}Observatorio Astron\'omico Nacional, Universidad Nacional de Colombia, Bogot\'a, Colombia.}
}

\maketitle

\begin{abstract}
Solutions of field equations in $f(R)$ gravity are found for a spherically symmetric and static spacetime in the Born-Infeld (BI) non-linear electrodynamics. It is found that the models supported in this configuration must have the parametric form $f'(R)|_{r}=m+n r$, with $m,n$ constants, whose value and sign have a strong impact on the solutions, as well as in the form and range of $f(R)$. When $n=0$, $f(R)=m R+m_0$ and the Einstein-BI solution is found. When $m\neq 0$ and $n\neq0$, $f(R)$ is asymptotically equivalent to GR and the Schwarzschild and $f(R)$-Reissner-Nordström solutions are written in some limits, likewise if $n>0$ and $r\gg1$, $f(R)$ can be found as a series approximation and as a particular case, when $R_S=-\frac{m^2}{3n}$, explicitly $f(R)=m R+2n\sqrt{R}+m_0$. Finally, the solutions, scalar curvature and parametric function $f(r)$ in the non-linear ($m=0$) regime of $f(R)$ are found, and some models for specific values of $m$ and $n$ are plotted.
\end{abstract}


Currently, there are two outstanding phenomena that cannot be explained from the neat gravitational interaction of baryonic matter in the frame of the General Relativity (GR) theory: the flat rotation curves of the galaxies at large radii \cite{zwicky} and the accelerated expansion of the Universe \cite{Riess:2004nr}. The approach to solve each problem was to consider that GR is indeed correct and assume the existence of a hypothetical, not interacting with radiation type of matter (Dark Matter [DM]) \cite{ostriker,Navarro_1996}, and a type of energy (Dark Energy [DE]) with repulsive gravity, represented in the Einstein field equations by the cosmological constant $\Lambda$ \cite{Peebles:2002gy}, respectively. The problem is that neither DM nor DE has yet been detected in the laboratory. Therefore a covariant theory of gravitation that does not depend on the cosmological scales and containing GR in some limit is needed.
\\
There are alternatives to DM model, like the Modified Newtonian Dynamics (MOND) theory, proposed by M. Milgrom \cite{Milgrom:1983ca}, where Newton's Laws are modified in order to reproduce the observations; or the polarization of the quantum vacuum, where DM is actually a dipolar fluid created by baryonic matter-antimmater pairs \cite{blanchet1,Hajdukovic_2011}; and alternatives to DE, like quintessence dynamic scalar field model \cite{ratra,Khurshudyan:2014yva} or extra spatial dimensions and brane cosmologies \cite{shiromizu,Dick_2001}, among others. 
\\
One of these alternatives is the $f(R)$ theory of gravity, which is of particular interest since it is constructed from a non-linear modification of the Einstein-Hilbert (EH) action without abandoning the structure of GR and instead being a natural extension of it. The main feature of $f(R)$ is to show the gravity as a more complex and intrinsic manifestation of the geometry-matter relationship, and at the same time it can be expressed as a scalar tensor theory with a new scalar degree of freedom \cite{kim,Faraoni:2010yi,Sotiriou:2008rp}, so it inherits the ability to pass all observational tests \cite{will2018}, and solves a wide range of cosmological problems like DM \cite{Boehmer:2007kx,Capozziello:2004km}, DE \cite{Capozziello:2004km,Nojiri:2006ri} or Inflation \cite{Starobinsky:1980te,Cognola:2007zu}.
\\
The new physics introduced by the non-linear terms of the curvature scalar can be traced in the solutions of the field equations, being the spherically symmetric and static (SSS) case the starting point. Different works have been done on this topic, including Schwarzschild like solutions of constant $R$ and $R=R(r)$ \cite{Capozziello_2008,Sebastiani_2011} and using the Noether symmetry approach \cite{capozziello2012}, charged SSS black hole solutions \cite{Nashed_2019}, and for particular models in cylindrically symmetric spaces \cite{Multam_ki_2006,MOMENI_2009}.

This work explores the solutions of field equations in SSS spaces in $f(R)$ gravity coupled with EM fields in the Born-Infeld theory\footnote{The Born-Infeld theory is a proposal of nonlinear electrodynamics with the objective of eliminating the singularity of electromagnetic fields present in the Maxwell's theory.} \cite{Born:1934gh}, and it is organized as follows: in section \ref{sec:2} the field equations are written under conditions of SSS. In section \ref{sec:GRvac} the GR-BI solution is shown as a particular case. The Schwarzschild-BI like and $f(R)$-Maxwell solutions are presented as limit cases of the $f(R)$-BI solution and the constraints on the forms of the models $f(R)$ generated by those solutions are shown in section \ref{sec:bif}. Section \ref{sec:binew} shows a particular solution of the $f(R)$-BI theory and its restrictions on the function $f(R)$. Conclusions are given in section \ref{sec:discu}.

\section{Field equations in Born-Infeld-f(R) theory}\label{sec:2}
The dynamics of the spacetime in $f(R)$ theory of gravity is determined by the field equations (\ref{fieldeq}), which are found from the least action principle applied to the action\footnote{Constructed from an arbitrary function, $f(R)$, of the scalar curvature, $R$, that generalizes the Einstein-Hilbert action, $\int d^4x \sqrt{-g}R$.} defined over a hypervolume $\Sigma$,
\begin{equation}\label{totalaction}
    I=\frac{1}{2\kappa}\left(\int_\Sigma d^4x \sqrt{-g}f(R)+I_{GYH}\right)+\int d^4x\sqrt{-g}L_M,
\end{equation}
where $I_{GYH}$, represents the Gibbons-York-Hawking term that corrects the boundary action value, see details in \cite{Guarnizo:2010xr}; and $L_M$ is the stress-energy lagrangian contribution, which in the BI formulation for a electromagnetic field in vacuum, is written as
\begin{equation}
    L_M(\mathrm{F})=b^2 \left(1-\sqrt{1+\frac{2 \mathrm{F}}{b^2}}\right),
\end{equation}
where $b$ is the BI parameter, so the fields in the Maxwell's theory turns out to be an approximation of the B-I theory in the region where the electromagnetic field is weak, or equivalently $b\to\infty$, while for $b\to0$, finite value of fields are obtained. At the same time, $\mathrm{F}=\frac{1}{4}F_{\mu\nu}F^{\mu\nu}$ is the Maxwell's classic lagrangian density, with the electromagnetic field tensor $F_{\mu\nu}=A_{\nu,\mu}-A_{\mu,\nu}$, and $A_\mu$ is the electromagnetic four-potential. Under spherical symmetry and staticity\footnote{In Gaussian units.} the only nonzero components of the field tensor are, without sources of magnetic fields,
\begin{equation}
    F_{t r}=-F_{r t}=\frac{q}{\sqrt{r^4+q^2/b^2}}.
\end{equation}
From (\ref{totalaction}), the field equations in the metric formalism, are found as
\begin{equation}\label{fieldeq}
    \mathcal{F}R_{\mu\nu}-\frac{1}{2}g_{\mu\nu}f-\mathcal{F}_{;\mu\nu}+g_{\mu\nu}\mathcal{F}_{;\alpha}^{;\alpha}=\kappa T_{\mu\nu},
\end{equation}
where $\mathcal{F}=f'(R)$, $f=f(R)$, and the geometry of the spacetime is described by the metric tensor $g_{\mu\nu}$, with $\mathcal{F}_{;\alpha}^{;\alpha}=g^{\alpha\beta}\left[f'(R)\right]_{;\alpha\beta}$, and
\begin{equation}\label{stressenergyBI}
    T_{\mu\nu}=g_{\mu \nu } L_M-F_{\mu \sigma } F_{\nu }^{\sigma }L_M',
\end{equation}
where the comma means total derivative with respect to $\mathrm{F}$. Note that unlike the Maxwell's theory, whose stress-energy tensor is traceless, tensor (\ref{stressenergyBI}) in the BI frame has $T=T_\mu^\mu\neq0$, which means that $R=R_0$ does not imply more $R_{\mu\nu}=\frac{1}{4}R_0g_{\mu\nu}$, and instead, by the trace equation
\begin{equation}
    f=\frac{1}{2}\left(\mathcal{F} R+3\mathcal{F}_{;\alpha }^{;\alpha }-\kappa T\right),
\end{equation}
field equations are reduced to
\begin{equation}\label{fieldequ}
    4\left(\mathcal{F} R_{\mu \nu }-\kappa  T_{\mu \nu }\right)-\left(\mathcal{F} R-\kappa T-\mathcal{F}_{;\alpha }^{;\alpha }\right)g_{\mu \nu }-4\mathcal{F}_{;\mu \nu }=0.
\end{equation}
This equation depends on the second covariant derivatives of the scalar function $f(R)$, which are combinations of partial derivatives of the metric, so to simplify the equations we will assume a spherically symmetric and static spacetime, defined by the metric
\begin{equation}
    ds^2=-a(r)dt^2+a^{-1}(r)dr^2+r^2d\theta^2+r^2 \sin^2\theta d\phi^2,
\end{equation}
from which the scalar curvature gives
\begin{equation}\label{scalarcurvature}
    R=\left[2-2a(r)-4r a'(r)\right]r^{-2}-a''(r),
\end{equation}
and the covariant derivatives of the function are found as
\begin{equation}
    \mathcal{F}_{;tt}=-\frac{1}{2}a(r)a'(r)\partial_r\mathcal{F}=a^2(r)\left(\partial^2_r\mathcal{F}-\mathcal{F}_{;rr}\right),
\end{equation}
\begin{equation}
    \mathcal{F}_{;\theta\theta}=ra(r)\partial_r\mathcal{F}=\csc^2\theta\mathcal{F}_{;\phi\phi},
\end{equation}
and
\begin{equation}
    \mathcal{F}_{;\alpha}^{;\alpha}=\left[a'(r)+\frac{2a(r)}{r}\right]\partial_r\mathcal{F}+a(r)\partial^2_r\mathcal{F}.
\end{equation}
In this way the components $tt$ and $rr$ of the field equations are (the other two components are indeed the same equation)
\begin{equation}
    4\frac{\mathcal{F} R_{t t}-\kappa  T_{t t}}{a(r)^2}+\frac{\mathcal{F} R-\kappa T}{a(r)}+\left[\frac{a'(r)}{a(r)}-\frac{2}{r}\right]\partial_r\mathcal{F}-\partial^2_r\mathcal{F}=0,
\end{equation}
\begin{equation}
    4\left(\mathcal{F} R_{r r}-\kappa  T_{r r}\right)-\frac{\mathcal{F} R-\kappa T}{a(r)}-\left[\frac{a'(r)}{a(r)}-\frac{2}{r}\right]\partial_r\mathcal{F}-3\partial^2_r\mathcal{F}=0,
\end{equation}
but since $R_{t t}=a^2(r)R_{r r}$ and $T_{t t}=a^2(r)T_{r r}$, adding the equations it must be fulfilled that
\begin{equation}
    \partial^2_r\mathcal{F}=0.
\end{equation}
this condition imposes a strong restriction on the form of the function $f(R)$, since it must satisfy
\begin{equation}
    \left.\frac{d f(R)}{dR}\right|_{R=R(r)}=m+nr,
\end{equation}
and by the fundamental theorem of calculus
\begin{equation}
    f(R)=\left[\int dr\left.f'(R)\right|_{R=R(r)}\partial_rR(r)\right]_{R(r)=R},
\end{equation}
or
\begin{equation}\label{condition}
    f(R)=\left[\int dr\left(m+n r\right) \partial_rR(r)\right]_{R(r)=R},
\end{equation}
with $m$ and $n$ constants. At first glance, one could try to find a solution $a(r)$ of the field equations by proposing a function $f(R)$ that satisfies it, simultaneously with definition of the scalar curvature (\ref{scalarcurvature}) and finally, if necessary, adjust the constants associated with the integrals by replacing $a(r)$ in the field equations. However proposing a viable model $f(R)$ that meets (\ref{condition}) and satisfies both the cosmological and the Solar system tests in the so called chameleon mechanism is not a simple task \cite{Brax_2008}, actually, the fact that the model satisfies the observational tests means that at some limit it is indistinguishable from the $\Lambda$-Cold Dark Matter model ($\Lambda$CDM) \cite{Perez_Romero_2018}. Despite this, the general solutions of the field equations will allows us to see the behavior of $R$ and thus restrict the functions $f(R)$ according to the constants $m$ and $n$, in three ways: (i) $m\neq0$ and $n=0$, (ii) $m\neq0$ and $n\neq0$ and (iii) $m=0$ and $n\neq0$.

\section{Einstein-BI solutions in vacuum}\label{sec:GRvac}
As expected, GR must be recovered when $f(R)=R$ or equivalently if $m\neq0$ and $n=0$, in which case the equations (\ref{fieldequ}) become
\begin{equation}\label{nzero}
4\left(m R_{\mu\nu}-\kappa T_{\mu\nu}\right)-\left(m R-\kappa T\right)g_{\mu\nu}=0,
\end{equation}
which are none other than Einstein field equations with $\kappa\to \kappa/m$
\begin{equation}
    r^2 a''(r)-2 a(r)-\frac{16 \pi  b q^2}{m \sqrt{q^2+b^2 r^4}}+2=0,
\end{equation}
whose solution is
\begin{equation}
    a(r)=1+c_2 r^2+\frac{c_1}{3 r}+\frac{8\pi b^2}{3 m}r^2\left[1-\sqrt{1+\frac{q^2}{b^2 r^4}}+\frac{2q^2}{b^2 r^4}\,_2F_1\left(\frac{1}{4},\frac{1}{2};\frac{5}{4};-\frac{q^2}{b^2 r^4}\right)\right],
\end{equation}
    
where $\,_2F_1$ is the Gaussian hypergeometric function and $c_1$ and $c_2$ are constant to be known at limit $b\to\infty$, that is
\begin{equation}
    a(r)=1+c_2 r^2+\frac{c_1}{3 r}+\frac{4\pi q^2}{mr^2},
\end{equation}
when $m=1$ the Reissner-Nordström (RN) solution must be recovered if $c_1=-3R_S$ and $c_2=0$, where $R_S$ is the Schwarzschild radius, and the RN-AdS space if $c_2=-\Lambda/3$. Then
\begin{equation}\label{sol1}
    a(r)=1-\frac{R_S}{r}-\frac{\Lambda}{3} r^2+\frac{8\pi b^2}{3 m}r^2\left[1-\sqrt{1+\frac{q^2}{b^2 r^4}}+\frac{2q^2}{b^2 r^4}\,_2F_1\left(\frac{1}{4},\frac{1}{2};\frac{5}{4};-\frac{q^2}{b^2 r^4}\right)\right],
\end{equation}
this solution is also obtained when $R=constant$ is assumed in the field equations, resulting $f'(R)=m$. Solution (\ref{sol1}) describes an electrically charged RN-AdS Black Hole solution in GR-BI theory \cite{hoffman,quevedo}, and the scalar curvature defines constant hypersurfaces
\begin{equation}
    R=\frac{16 \pi b^2}{m} \left[\left(2+\frac{q^2}{b^2 r^4}\right)\left(1+\frac{q^2}{b^2 r^4}\right)^{-1/2}+\frac{m\Lambda}{4\pi b^2}-2\right],
\end{equation}
such that $R_{b\to0,\infty}=4\Lambda$ as it must be. Now, by (\ref{condition}), the function can be obtained directly
\begin{equation}
    f(R)=m R+m_0,
\end{equation}
where $m_0$ is some constant, i.e. the models are reduced to the $\Lambda$-CDM model with $\kappa\to\kappa/m$.
\section{$f(R)$-BI solutions}\label{sec:bif}
In the most general case, when $m\neq0$ and $n\neq0$, the field equations take the form
\begin{equation}
    r^2\left[(m+n r)a''(r)+n a'(r)\right]-2a(r)(m+2n r)+2(m+n r)=\frac{16 \pi  b q^2}{\sqrt{b^2 r^4+q^2}}
\end{equation}
although the solution cannot be found analytically but expressed in terms of an integral, as follows
\begin{equation}\label{generalsolution}
     a(r)=1+\frac{R_S}{m}\left(\frac{3n}{2m}-\frac{1}{r}\right)+\frac{n}{m}\left(1+\frac{3n R_S}{m^2}\right)\left(\ln\left[\frac{m}{r}+n\right]^{\frac{n}{m}}-\frac{1}{r}\right)r^2-16 \pi  q^2 r^2 \int dr \frac{\, _2F_1\left(\frac{1}{4},\frac{1}{2};\frac{5}{4};-\frac{q^2}{b^2 r^4}\right)}{r^5 (m+n r)}
\end{equation}
it is still possible to obtain information about the form of the functions because the logarithm must be defined in the following intervals depending on the signs of the constants
\begin{itemize}
    \item $I_1$: $r>0$ if $m,n>0$,
    \item $I_2$: $r>|\frac{m}{n}|$ if $m<0$ and $n>0$,
    \item $I_3$: $0<r<|\frac{m}{n}|$ if $m>0$ and $n<0$.
\end{itemize}
We are modeling the stress-energy tensor with the electromagnetic lagrangian density in the Born-Infeld theory, so there are two natural cases of physical interest, $b=0$ (or equivalently $q=0$): $f(R)$ gravity in a SSS vacuum spacetime without the presence of electromagnetic fields. And $b\to\infty$: $f(R)$ gravity coupled with classic electromagnetic fields, which are described below.

\subsection{$f(R)$-Schwarzschild-like solution}
\begin{figure}[ht]
\centering
\subfigure[$n=1.2$.]{
\resizebox*{8cm}{!}{\includegraphics{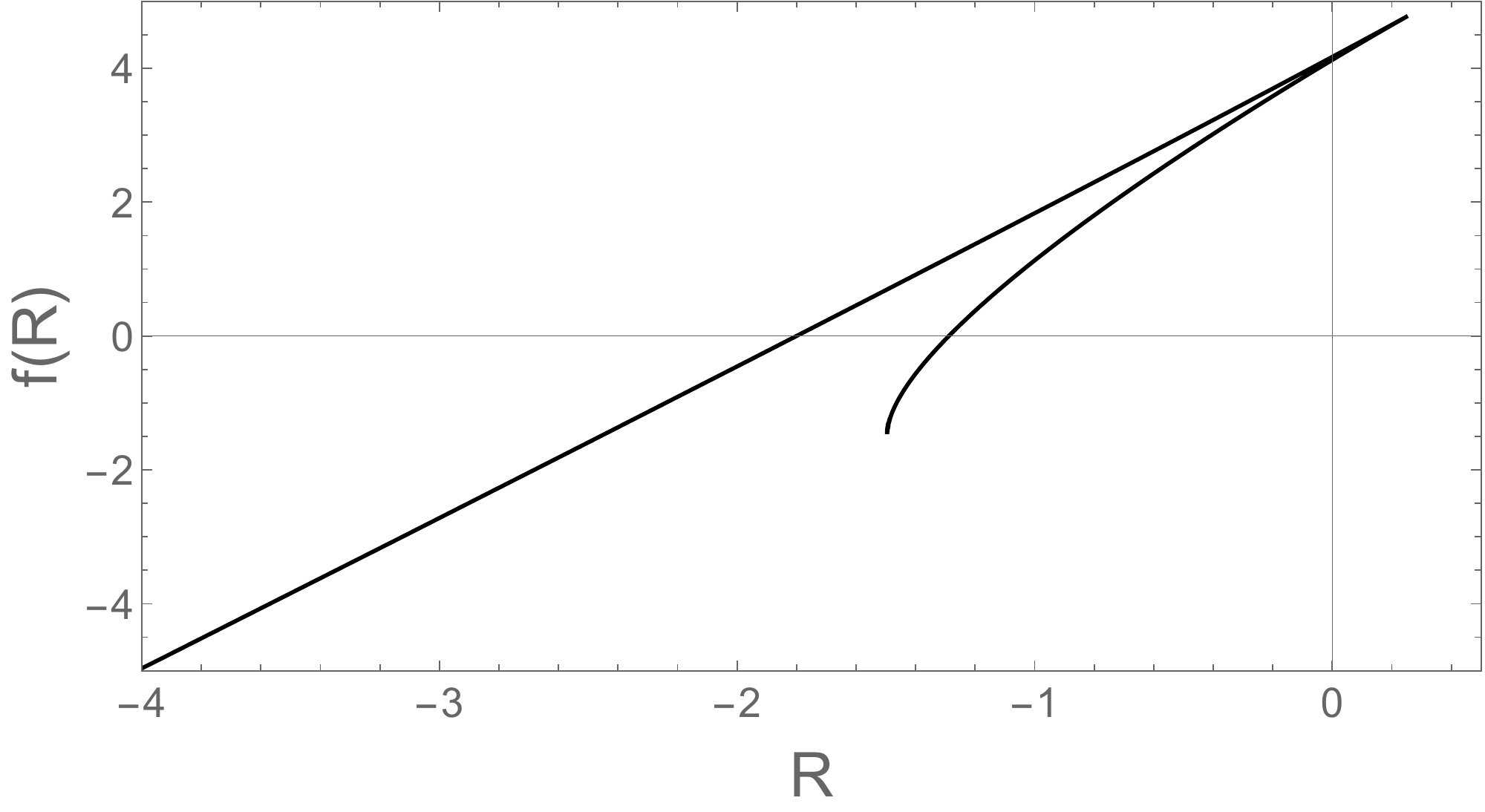}}}\hspace{5pt}
\subfigure[$n=-1.383,-1.333,-1.283$ (dashed, continuous, dotted).]{
\resizebox*{8.2cm}{!}{\includegraphics{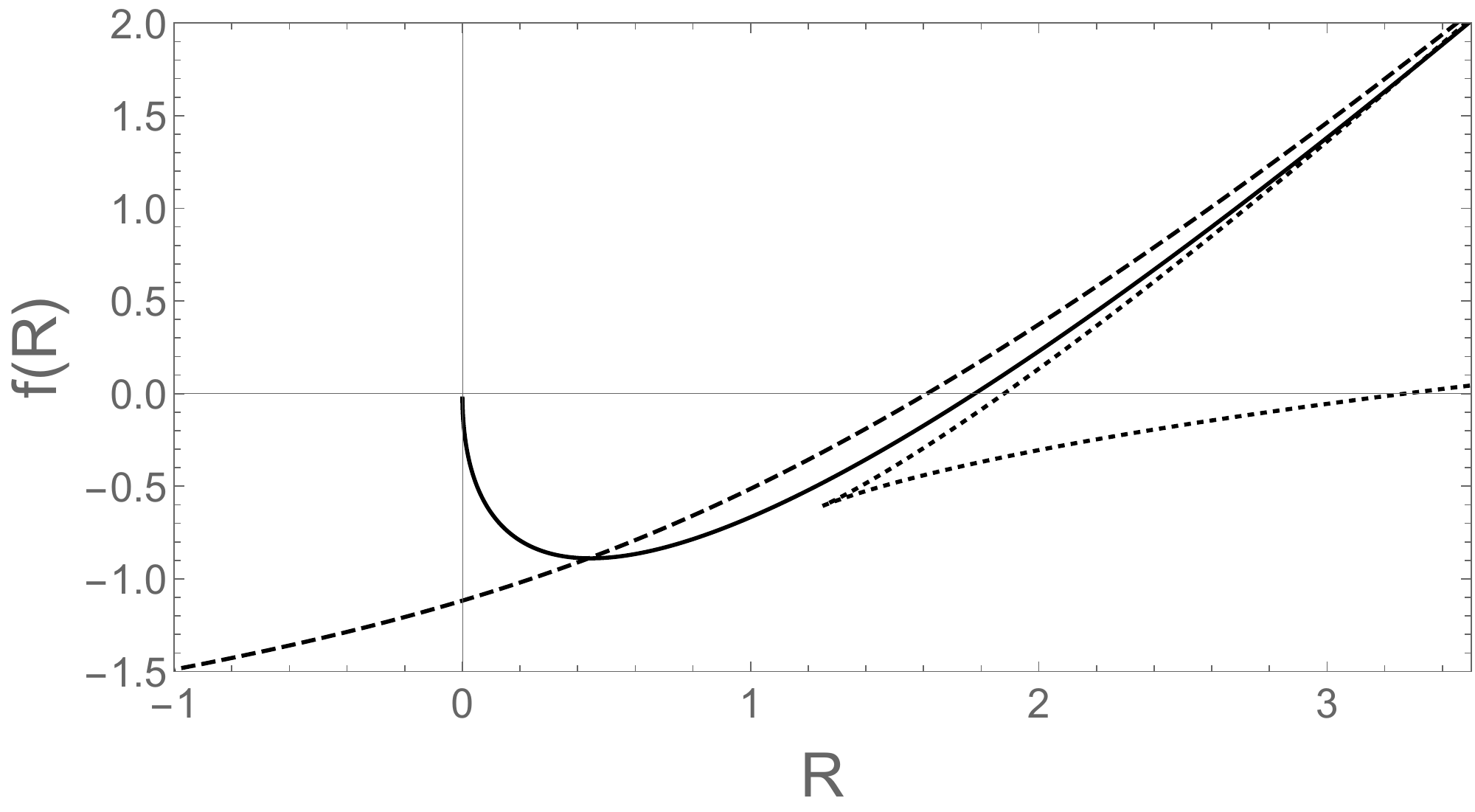}}}
\caption{Models $f(R)$ allowed for the Schwarzschild-like space and some values of $m$ and $n$. In (a), when $r_c=0.397$, $R=0.253$ and $f(R)=4.780$, which is the absolute maximum, and the spatial infinity is mapped to the point (-1.496,-1.496). In (b), the three possible sub-cases depending on $h$ Eq. (\ref{h}). In both panels $m=2$ and $R_S=1$.}\label{fig1}
\end{figure}

In the first case ($b=0$) the solution is
\begin{equation}\label{solution1}
    a(r)=1+\frac{R_S}{m}\left(\frac{3n}{2m}-\frac{1}{r}\right)+\frac{n}{m}\left(1+\frac{3n R_S}{m^2}\right)\left(\ln\left[\frac{m}{r}+n\right]^{\frac{n}{m}}-1\right)r^2,
\end{equation}
this correspond to the Schwarzschild-like solution in $f(R)$ theory for a non-constant scalar curvature, which gives
\begin{equation}\label{scalarb0}
    R=\frac{m n}{(m+n r)^2}\left[\frac{6}{r}+\frac{19 n}{m}+\frac{12n^2}{m^2}r-\frac{3R_S}{m} \left(\frac{2n}{m}+\frac{1}{r}\right) \left(\frac{1}{r}-\frac{6n}{m}-\frac{6n^2}{m^2}r\right)\right]-\left(1+\frac{3 n R_S}{m^2}\right) \ln\left[\frac{m}{r}+n\right]^{\frac{12n^2}{m^2}},
\end{equation}
and the function
\begin{equation}
    f(r)=\frac{n}{m+n r} \left[\frac{6m}{r}+8n-9Rs \left(\frac{1}{3r^2}-\frac{n}{m r}-\frac{2n^2}{m^2}\right)\right]-\left(1+\frac{3n R_S}{m^2}\right) \ln\left[n+\frac{m}{r}\right]^{\frac{6n^2}{m}}+m_0.
\end{equation}
scalar (\ref{scalarb0}) cannot be analytically inverted for all $m$ and $n$, so it is not possible to express the function $f$ in terms of $R$, moreover it presents singularities at the limits of the regions mentioned above, $r\to0$ in $I_1$, $r\to|\frac{m}{n}|$ in $I_2$ and the combination of the previous two in $I_3$. Notwithstanding the foregoing, we can qualitatively analyze the behavior of the scalar with respect to $r$ in each of the three regions and draw $f(R)$ by numerically inverting Eq. (\ref{scalarb0}) when assigning values to $m$ and $n$ according to each of the regions. For this purpose we must consider the critical points of $R(r)$ and $f(R)$ since these will determine the domain $R$ in which $f(R)$ can be defined. Note that by Eq. (\ref{condition})
\begin{equation}
    \partial_rf(r)-(m+nr)\partial_r R(r)=0,
\end{equation}
both functions share the same maximum points. If case $I_1$ or $I_3$ (if $n>-\frac{m^2}{3R_S}$) there is one absolute maximum and minimum, respectively at
\begin{equation}\label{critical}
    r_{c}=\frac{m}{n}\left(h^{1/3}-1\right),
\end{equation}
of value
\begin{equation}\label{rc}
    R_c=n \frac{1-h}{R_S}\left(\frac{2}{3}+\frac{1-4h+2h^{2/3}}{h^{1/3}-1}+4h\ln \left[\frac{n h^{1/3}}{h^{1/3}-1}\right]\right),
\end{equation}
where 
\begin{equation}\label{h}
    h=1+\frac{3R_S n}{m^2}.
\end{equation}
which is precisely the term accompanying the logarithm. In $I_1$ for $0<r<r_c$ $f(R)$ will be an increasing function, and in $r=r_c$ the function folds and when $r\to\infty$ $R(r)$ and $f(r)$ take the constant values 
\begin{equation}\label{limit}
    \left.R\right|_{r\to\infty}=R_\infty=-\frac{12n^2}{m^2}h\ln n,
\end{equation}
valid for $I_1$ and $I_2$ since $r$ is bounded in $I_3$, and
\begin{equation}\label{limit2}
    \left.f(R)\right|_{r\to\infty}=\frac{m}{2}R,
\end{equation}
which is valid not only for $I_1$, but in all cases, so the solution of GR is recovered in the spatial infinity. However, at $r\to\infty$ the series expansion of $R(r)$ and $f(r)$ produces, for $n>0$
\begin{equation}
    R(r)=\left.R\right|_{r\to\infty}+\frac{1}{r^2}+O\left(\frac{1}{r}\right)^5,
\end{equation}
and
\begin{equation}
    f(r)=\left.f(R)\right|_{r\to\infty}+\frac{2n}{r}+\frac{m}{r^2}-\frac{hm^3}{2n^2r^4}+O\left(\frac{1}{r}\right)^5,
\end{equation}
thus for large $r$, function can be approximated to
\begin{equation}\label{approx}
    f(R)\approx m\left(R-\frac{1}{2}R_{\infty}\right)+2n\sqrt{R-R_{\infty}}-\frac{hm^3}{2n^2}\left(R-R_{\infty}\right)^2.
\end{equation}
In Fig. (\ref{fig1}) (a) it is shown $f(R)$ for some values of the constants according to $I_1$, where it should be noted that as $r$ tends to $r_c$ in the interval $(0,r_c)$, the behavior of $f(R)$ is almost linear. In case $I_2$, $f(r)$ and $R(r)$ are decreasing and do not present maximum or minimum, so that $f(R)$ is a monotonically increasing function whose domain is $R>\left.R\right|_{r\to\infty}$. In case $I_3$, if $h>0$, functions $f(r)$ and $R(r)$ have an absolute minimum at $r_c$, Eq. (\ref{critical}), and $f(R)$ will present a fold, Fig. (\ref{fig1}) (b), while for $h<0$ functions are monotonically decreasing and since Eq. (\ref{limit}) is defined for $n>0$, the domain of $f(R)$ is $(-\infty,\infty)$, in such a way that Eq. (\ref{limit2}) is fulfilled and
\begin{equation}
    \left.\frac{f(R)}{R}\right|_{r\to0}=m,
\end{equation}
which implies that the solutions are asymptotically similar to GR.
However there is a third sub-case of special interest because the scalar is invertible and therefore the function $f(R)$ can be expressed directly. This is the case when $h=0$, the logarithmic term vanishes and the solution (\ref{solution1}) is simplified to
\begin{equation}
    a(r)=\frac{1}{2}+\frac{m}{3n r},
\end{equation}
the scalar
\begin{equation}\label{scalarsimple}
    R=\frac{1}{r^2},
\end{equation}
and by (\ref{condition}), the function could only be
\begin{equation}\label{functionsimple}
    f(R)=m R+2n\sqrt{R}+m_0,
\end{equation}
Eq. (\ref{approx}) reproduces this function when $h=0$ and its behaviour can be seen in Fig. (\ref{fig1}) (b) for $m_0=0$, $m=2$, $n=-4/3$ and $R_S=1$. Model (\ref{functionsimple}) can be interpreted as a a perturbation of the Ricci scalar in GR around the vacuum solution, moreover, it is a particular case ($m_0=0$, $m=1$, $n=-\alpha/2$) of the family models for DE \cite{Amendola_2007b,Amendola_2007a,carrollfr}
\begin{equation}
    f(R)=R-\alpha^{2(\beta+1)}R^{-\beta},
\end{equation}
with $\beta=1/2$. Likewise, the Black Hole thermodynamic properties and stability in this models are studied in \cite{Elizalde_2020}. Note that (\ref{functionsimple}) is the first order expansion at $R=0$ of
\begin{equation}\label{functionappr}
    f(R)=\frac{2n}{\bar{n}\left(1-\bar{n}\sqrt{R}\right)},
\end{equation}
when $m_0=\frac{4 n^2}{m}$ and $\bar{n}=\frac{m}{2n}$, moreover Eq. (\ref{functionappr}) is a good approximation of model (\ref{functionsimple}) for $|n|\gg|m|$.

\subsection{$f(R)$-RN solution}
The second case of interest, which actually contains the Schwarzschild solution, is when $b\to\infty$, i.e. $f(R)$-Maxwell theory for a SSS spacetime, and the solution is given by
\begin{multline}\label{secondsolu}
    a(r)=1+\frac{n}{m}\left[\frac{3R_S}{2 m}+\frac{4 \pi q^2}{n} \left(\frac{2 n^2}{m^2}+\frac{1}{r^2}\right)-\left(\frac{16 \pi  q^2}{3 m}+\frac{R_S}{n}\right)\frac{1}{r}+\left(1+\frac{3 n R_S}{m^2}+\frac{16 \pi  n^2 q^2}{m^3}\right)\right.\\
    \left.\left(\ln\left[\frac{m}{r}+n\right]^{\frac{n}{m}}-\frac{1}{r}\right)r^2\right],
\end{multline}
the scalar curvature
\begin{multline}\label{scalarbinfty}
    R=\frac{2 n^2}{(m+n r)^2} \left\{\frac{19}{2}+\frac{27 n R_S}{m^2}+\frac{144 \pi  n^2 q^2}{m^3}-\left(\frac{3 R_S}{2 n}+\frac{8 \pi  q^2}{m}\right)\frac{1}{r^2}+\frac{2 m}{n} \left(\frac{3}{2}+\frac{3 n R_S}{m^2}+\frac{16 \pi  n^2 q^2}{m^3}\right)\frac{1}{r}+\right.\\
    \left.\frac{6 n}{m} \left(1+\frac{3 n R_S}{m^2}+\frac{16 \pi  n^2 q^2}{m^3}\right) \left(r-\left(\frac{m}{n}+r\right)^2 \ln\left[\frac{m}{r}+n\right]^{\frac{n}{m}}\right)\right\}\;,
\end{multline}
and the function is determined by
\begin{multline}
    f(r)=\frac{6 n^2}{m+n r}\left[\frac{4}{3}+\frac{3 n R_S}{m^2}+\frac{16 \pi  n^2 q^2}{m^3}-\left(\frac{R_S}{2 n}+\frac{8 \pi  q^2}{3 m}\right)\frac{1}{r^2}+\left(\frac{m}{n}+\frac{3 R_S}{2 m}+\frac{8 \pi  n q^2}{m^2}\right)\frac{1}{r}\right]-\\
    \left(1+\frac{3 n R_S}{m^2}+\frac{16 \pi  n^2 q^2}{m^3}\right)\ln\left[\frac{m}{r}+n\right]^{\frac{6 n^2}{m}},
\end{multline}
from where it is recognized the term that modulates the logarithm
\begin{equation}\label{hbar}
    \bar h=1+\frac{3n R_S}{m^2}+\frac{16\pi n^2 q^2}{m^3},
\end{equation}
in this sense, the classification that was made according to the value of the constants is also useful to find the critical points, and Eq. (\ref{critical}) remains valid, as well as the limit (\ref{limit}) but with $h\to\bar h$, such that Eq. (\ref{limit2}) and (\ref{approx}) are also fulfilled. Note that the last term of $\bar h$ is quadratic in $q$ and $n$, so it does not affect the number of critical points when $m>0$ and the analysis made in the previous section is still valid for $I_1$, however in $I_2$, when $\bar h<0$, there will be an absolute maximum at
\begin{equation}
    r_c=-\frac{m}{n}\left[(-\bar h)^{1/3}+1\right],
\end{equation}
this critical point will produce a bend in the plot $f(R)$ at $R_c$, therefore its domain will be $R\leq R_c$. At the other hand, if $\bar h>0$, $R(r)$ and $f(r)$ are monotonically decreasing, so $f(R)$ will be increasing with domain $R>R|_{r\to\infty}$. In $I_3$ the appearance of the critical point occurs when $\bar h<1$, that is when $m<\frac{9R_S^2}{64\pi q^2}$ and at the same time
\begin{subequations}
\label{ineq1}
\begin{equation}
    n\geq\frac{m}{32\pi q^2}\left(\sqrt{9R_S^2-64m\pi q^2}-3R_S\right)\quad\text{or}\quad-\frac{3m}{16\pi q^2}<n\leq-\frac{m}{32\pi q^2}\left(\sqrt{9R_S^2-64m\pi q^2}+3R_S\right),
\end{equation}
\end{subequations}
or when
\begin{equation}\label{ineq2}
    m\geq\frac{9R_S}{64\pi q^2}\qquad\text{and}\qquad n\geq-\frac{3m R_S}{16\pi q^2}.
\end{equation}
In thoses cases, $f(R)$ will be defined for $R_c\leq R$, otherwise the domain of $f(R)$ will be $R<R|_{r\to\infty}$.

\begin{figure}[ht]
\centering
\subfigure[$\bar h=0, 1, -1/2, -1/18$, Eq. (\ref{hbar}) (continuous, dotted, dashed and dotdashed lines) for $m=-3$, $n=1$ and $q=1$.]{
\resizebox*{8.2cm}{!}{\includegraphics{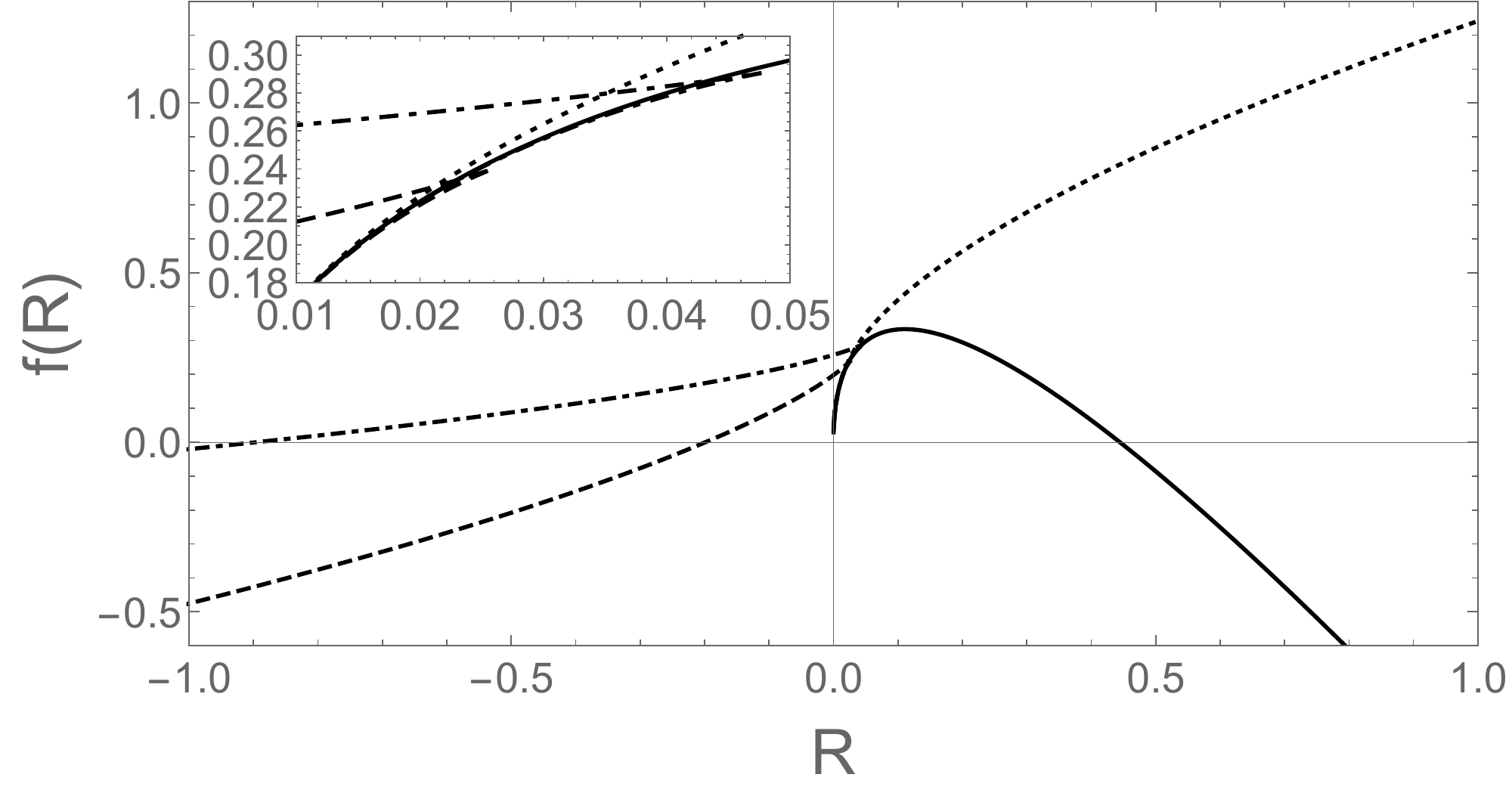}}}\hspace{5pt}
\subfigure[$b=0$ ($q=0$) and $n=-1,1$ (dotdashed and dotted lines), and $b\to\infty$ ($q=1$) for $n=-1,1$ (continuous and dashed lines).]{
\resizebox*{8cm}{!}{\includegraphics{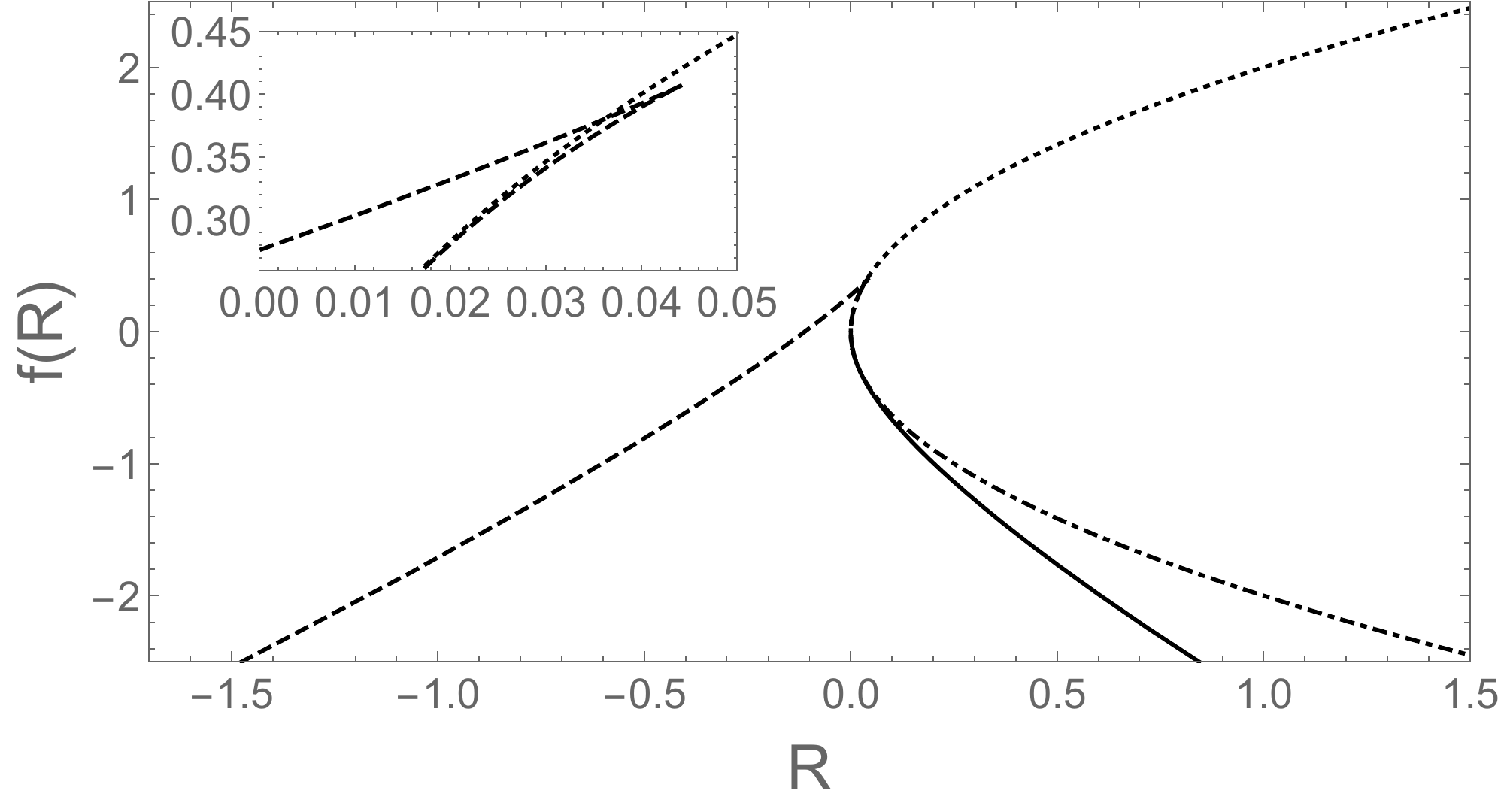}}}
\caption{Functions $f(R)$ found numerically for some values of $m$ and $n$ according to the $f(R)$-RN solution (\ref{secondsolu}), (a), and $f(R)$-non-linear-BI (\ref{solutionn}), (b). The convergence of the functions is observed when $r$ goes from the maximal point (if any) to infinity, or equivalently for $0<R<R_c$, this fact is explained by Eq. (\ref{approx}). In (a) each curve has a different Schwarzschild radius. Although the solutions are not equal because they depend on the presence ($b\to\infty$) or not ($b\to0$) of the Maxwell fields as well as the $m$ and $n$ constants, dotted and dashed lines in both panels have similar form, moreover $f(R)$ represented by dashed lines do not have continuous derivative in $R_c$} \label{fig2}
\end{figure}

When observing the scalar (\ref{scalarbinfty}), four posibilities of choosing the constants when $n>0$ are highlighted and their importance consists in the form and simplicity that the solution takes and correspondingly the function $f(R)$.
\begin{enumerate}
    \item If $\bar h=0$, the solution does not depend on the $r$ and $r^2$ terms, including the logarithmic one,
\begin{equation}
    a(r)=\frac{1}{2}+\frac{4 \pi  q^2}{m r^2}+\frac{m}{3 n r},
\end{equation}
the scalar is just eq. (\ref{scalarsimple}) and the function is given by eq. (\ref{functionsimple}), this fact is explained because any solution of the form $a(r)=\frac{1}{2}+\frac{\alpha}{r}+\frac{\beta}{r^2}$, produces $R=\frac{1}{r^2}$, with $\alpha$ and $\beta$ some constants.
\item If $\bar h=1$, the term $r$ disappears in the solution, which is given by
\begin{equation}
    a(r)=1+\frac{4 \pi  q^2}{m r^2}-\frac{n r}{m}+\frac{n^2 r^2}{m^2} \ln \left[\frac{m}{r}+n\right],
\end{equation}
the scalar and the function are respectively
\begin{equation}
    R=\frac{6 n}{m r}+\frac{n^2}{(m+n r)^2} \left(7+\frac{6 n}{m}r\right)-\frac{12 n^2}{m^2} \ln \left[\frac{m}{r}+n\right],
\end{equation}
\begin{equation}
    f(r)=2 n \left(\frac{3}{r}+\frac{n}{m+n r}-\frac{3 n}{m} \ln \left[\frac{m}{r}+n\right]\right).
\end{equation}
\item If $\bar h=-\frac{1}{2}$, the solution is
\begin{equation}\label{solution3}
    a(r)=\frac{1}{4}+\frac{4 \pi  q^2}{m r^2}+\frac{m}{2 n r}+\frac{n}{2 m}r-\frac{n^2}{2 m^2} r^2\ln \left[\frac{m}{r}+n\right],
\end{equation}
but the scalar does not depend on $r^{-1}$ term
\begin{equation}\label{scalar3}
    R=-\frac{n^2}{(m+n r)^2} \left(8-\frac{3 m^2}{2 n^2 r^2}+\frac{6 n r}{m}\right)+\frac{6 n^2}{m^2}\ln \left[\frac{m}{r}+n\right],
\end{equation}
and the function
\begin{equation}
    f(r)=\frac{3 m}{2 r^2}-\frac{n^2}{m+n r}+\frac{3 n^2}{m} \ln \left[\frac{m}{r}+n\right].
\end{equation}
\item If $\bar h=-\frac{1}{18}$, the solution has a similar form to the previous Eq. (\ref{solution3})
\begin{equation}
    a(r)=\frac{17}{36}+\frac{4 \pi  q^2}{m r^2}+\frac{19 m}{54 n r}+\frac{n r}{18 m}-\frac{n^2 r^2}{18 m^2} \ln \left[\frac{m}{r}+n\right],
\end{equation}
however the scalar contains a $r^{-1}$ term that (\ref{scalar3}) does not have
\begin{equation}
    R=\frac{m^2}{3(m+n r)^2} \left(\frac{19}{6 r^2}+\frac{16 n}{3 m r}-\frac{2 n^3 r}{m^3}\right)+\frac{2 n^2}{3 m^2} \ln \left[\frac{m}{r}+n\right],
\end{equation}
and function
\begin{equation}
    f(r)=\frac{1}{9} \left(\frac{19 m}{2 r^2}+\frac{16 n}{r}-\frac{n^2}{m+n r}\right)+\frac{3 n^2}{m} \ln \left[\frac{m}{r}+n\right].
\end{equation}
\end{enumerate}
Note that the value of $R_S$ is determined by constants $m$ and $n$ according of each case, however, because the limit (\ref{limit2}) only depends on $m$, functions $f(R)$ are expected to coincide when $r\to\infty$. This can be seen in Fig. (\ref{fig2}) (a), where $n$ was established from $\left.R\right|_{r\to\infty}$, since although this limit depends on $\bar h$, it also depends on $n$, so when $n=1$, the limits coincide.

\section{$f(R)$-non-linear-BI model}\label{sec:binew}
When $m=0$ and $n\neq0$ the models $f(R)$ do not have the linear term $R$, and thus are separated from GR. the solution cannot be obtained directly, as might be supposed, from Eq. (\ref{generalsolution}) in the most general case, so it is necessary to write the field equations,
\begin{equation}
    r^2\left[r a''(r)+a'(r)\right]-4 r a(r)-\frac{16 \pi  b q^2}{n\sqrt{b^2 r^4+q^2}}+2 r=0,
\end{equation}
with
\begin{equation}\label{solutionn}
    a(r)=\frac{1}{2}+c_2 r^2+\frac{c_1}{4 n r^2}+\frac{16 \pi  q^2}{5 n r^3} \, _2F_1\left(\frac{1}{4},\frac{1}{2};\frac{9}{4};-\frac{q^2}{b^2 r^4}\right),
\end{equation}
where $c_1$ and $c_2$ are constant to be determined in some limit. However the curvature scalar does not depend on $c_1$
\begin{equation}
    R=-12 c_2+\frac{1}{r^2}+\frac{16 \pi b^2}{n r}\left[\left(1+\frac{q^2}{b^2 r^4}\right)^{-1/2}-\, _2F_1\left(\frac{1}{4},\frac{1}{2};\frac{5}{4};-\frac{q^2}{b^2 r^4}\right)\right],
\end{equation}
with which the function is
\begin{equation}
    f(r)=\frac{2n}{r}+\frac{16\pi b^3 r^2}{\sqrt{b^2 r^4+q^2}}.
\end{equation}
however when $r\gg1$,
\begin{equation}
    R(r)=-12c_2+\frac{1}{r^2}+O\left(\frac{1}{r}\right)^5,
\end{equation}
and
\begin{equation}
    f(r)=16\pi b^2+\frac{2n}{r}-\frac{8\pi q^2}{r^4}+O\left(\frac{1}{r}\right)^5
\end{equation}
thus the function can be approximated to
\begin{equation}\label{approx2}
    f(R)\approx16\pi b^2+2n\sqrt{12c_2+R}-8\pi q^2(12c_2+R)^2,
\end{equation}
this function has a similar form to Eq. (\ref{approx}) with exception to the linear term $R$, both expressions contain the $2n\sqrt{R}$ term, but independent of $c_2$, $m$ must not be real for the $R^2$ terms to match. In addition, $f(R)$ (\ref{approx2}) is only defined at limit $b=0$ or SSS vacuum space without Maxwell fields ($q=0$), in this case the solution is reduced to
\begin{equation}
    a(r)=\frac{1}{2}+\frac{c_1}{4n r^2}+c_2 r^2,    
\end{equation}
the scalar is
\begin{equation}
    R=-12 c_2+\frac{1}{r^2},
\end{equation}
and function
\begin{equation}
    f(R)=m_0+2n\sqrt{12c_2+R}.
\end{equation}
where $m_0$ is some constant. At the other hand, $b\to\infty$, leads to the solution
\begin{equation}
    a(r)=\frac{1}{2}+\frac{16\pi q^2}{5n r^3}+\frac{c_1}{4n r^2}+c_2r^2,
\end{equation}
with the scalar
\begin{equation}
    R=-12c_2+\frac{1}{r^2}-\frac{32\pi q^2}{5n r^5},
\end{equation}
and the function
\begin{equation}
    f(r)=-\frac{8\pi q^2}{r^4}+\frac{2n}{r}.
\end{equation}
Since $m=0$, the scalar curvature as well as the function $f(R)$ will be determined by the sign of $n$ as can be seen in Fig. (\ref{fig2}) (b) where are plotted the 4 possible forms of the functions. When $n>0$ $R(r)$ has a similar form to the $f(R)$-RN models when $\bar h=1,-1/2$, and for Maxwell fields $R(r)$ has a maximal point and $f(R)$ a corresponding fold at $R_c$ while for $n<0$ the function will be monotonically decreasing.

\section{Concluding remarks}\label{sec:discu}
We have studied in this work the solutions of the field equations in spherically symmetric and static spaces in the $f(R)$ theory of gravity in the presence of a non-linear electromagnetic fields in the BI frame, depending on the parameter $b$, which in turn exhibits two limits of classical physical meaning, $f(R)$ Schwarzschild-like ($b\to0$) and $f(R)$ RN ($b\to\infty$) solutions. We found that the only models allowed in this framework must have the parametric form, $\mathcal{F}=m+n r$, with $m, n$ constants. From this condition it is possible to determine the form, domain and range of the models $f(R)$ allowed, and a classification of the solutions based on the values of the constants allows to find different solutions.
When $n=0$, GR is recovered with the rescaling $\kappa\to\kappa/m$ and $f(R)=m R+m_0$.

When $m\neq0$ and $n\neq0$ $f(R)$ is asymptotically equivalent to GR, and although the solution cannot be written analytically, writing it at the limits mentioned above and by subclassing the constants according to their sign, it is possible to analyze the functions. In the $f(R)$-Schwarzschild case the solution, Ricci scalar and the parametric function are found in terms of $r$, besides if
\begin{itemize}
    \item $n,m>0$: $f(R)$ is defined for $R<R_c$, where $R_c$ is given by (\ref{rc}), however it has no continuous derivative and thus does not represent a viable physical model.
    \item $m<0$ and $n>0$: $f(R)$ is monotonically increasing with $R>-\frac{12 n^2}{m^2}h \ln n$.
    \item $m>0$ and $n<0$: when $-3n R_S<m^2$, $f(R)$ has no continuous derivative and $R>R_c$. When $-3n R_S>m^2$, $f(R)$ is monotonically increasing for all $R\in\Re$. When $-3n R_S=m^2$, $f(R)=m R+2n\sqrt{R}+m_0$, which is a model for the expansion of the Universe without DE.
\end{itemize}
For $f(R)$-RN case, being a generalization of the previous one with electric charge, the above arguments remain valid, but with $\bar h$ (\ref{hbar}). However the existence of $q$ means that the function $f(R)$ has no longer continuous derivative for all $m<0$ and $n>0$, and thus defined only for $R\leq R_c$ or $R>-\frac{12 n^2}{m^2}\bar h\log n$, or vice versa in the opposite case, $m>0$ and $n<0$. When $n>0$ and depending on $\bar h$, different solutions are found, which affects the form of the functions, although for $r\gg1$ all these coincide according to Eq. (\ref{approx}).

When $m=0$ we depart from the lineal term of GR, although this does not mean that the model has to be discarded, on the contrary, the curves of the models show that there are indeed similarities with some functions in the $f(R)$-RN solution and as a mathematical toy is valid. In this case the solution is written in terms of a hypergeometric function, and since it is not possible to solve $f$ in terms of $R$ for all $r$, when $r\gg1$ $f(R)$ can be given as a approximation, Eq. (\ref{approx2}), which is linearly independent to Eq. (\ref{approx}) and thus representing different models.

\bibliography{biblio}

\end{document}